\documentclass{article}

\usepackage{graphicx}

\newtheorem{thm}{Theorem}

\newtheorem{fig}{Figure}

\def\leurre{\noindent\leftskip0pt\small\baselineskip 10pt}

\def\encadre#1#2{%
\setbox100=\hbox{\kern#1{#2}\kern#1}
\dimen100=\ht100 \advance \dimen100 by #1
\dimen101=\dp100 \advance \dimen101 by #1
\setbox100=\hbox{\vrule height \dimen100 depth \dimen101\box100\vrule}
\setbox100=\vbox{\hrule\box100\hrule}
\advance \dimen100 by .4pt \ht100=\dimen100
\advance \dimen101 by .4pt \dp100=\dimen101
\box100
\relax
}

\def\ligne#1{\hbox to \hsize{#1}}
\def\PlacerEn#1 #2 #3 {\rlap{\kern#1\raise#2\hbox{#3}}}

\setbox221=\hbox{\includegraphics{cercle_L20.eps}}

\def\encercle#1#2{\hbox{\raise-5pt\copy221\hskip#2#1}}

\begin{document}

\title{Infinigons of the hyperbolic plane and grossone}

\author{Maurice {\sc Margenstern}\\
\small
professor emeritus,\\
Universit\'e de Lorraine,\\
Laboratoire d'Informatique Th\'eorique et Appliqu\'ee, EA 3097,\\
\small
Campus de Metz,\\
\small
\^Ile du Saulcy,\\
\small
57045 Metz Cedex, France,\\
\small
{\it email}\ {margenstern@gmail.com}\\
\small
{\tt http://www.lita.sciences.univ-metz.fr/~{}margens}\\
\\
\small
}
\maketitle
\begin{abstract}
In this paper, we study the contribution of the theory of grossone to the study
of infinigons in the hyperbolic plane. We can see that the theory of grossone
can help us to obtain a much more classification for these objects than in the
traditional setting.
\end{abstract}

ACM-class: 
F.2.2., F.4.1, I.3.5

{\bf keywords}:
tilings, hyperbolic plane, infinigons, grossone
\vskip 10pt

\def\cqfd{\hbox{\kern 2pt\vrule height 6pt depth 2pt width 8pt\kern 1pt}}
\def\Hii{$I\!\!H^2$}
\def\Hiii{$I\!\!H^3$}
\def\Hiv{$I\!\!H^4$}
\def\norm{\hbox{$\vert\vert$}}

   In~\cite{mm_infigFI}, an algorithmic approach to the infinigons was given by this
author.

   Infinigons of the hyperbolic plane are polygons with infinitely many sides. It is the case
that there are infinitely many such objects and that, among them, there is an infinite family
which tiles the hyperbolic plane by applying to an initial infinigon the process which is used
to obtain a tessellation from an ordinary regular convex polygon of that plane.
The existence of infinigons which tiles the plane appear already in~\cite{cox_infig} and
in~\cite{rozen_infig}. In~\cite{mm_infigFI}, it was proved that for any angle~$\alpha$
with $\alpha\in]0..\pi[$ it is possible to construct an infinigon such that consecutive
sides make an angle of~$\alpha$. Moreover, such an
infinigon tiles the plane by reflection in its sides and, recursively, of its images in their 
sides, when $\alpha=\displaystyle{{2\pi}\over k}$ with $k$ being a positive integer with
$k\geq3$ and only in this case. As already mentioned, \cite{mm_infigFI} gives an algorithmic
construction for the tiling defined by an infinigon whose angle is $\displaystyle{{2\pi}\over k}$
with $k\geq3$.

   Of course, when we speak of an infinite object or infinitely many objects in the framework
of grossone, it comes to our mind that we have to make use of more precise terms. When
we speak of an infinite family of infinigons, we have of course to make more precise
how infinite our family is and how infinite our infinigons themselves are.

   In Section~\ref{clas_infig}, we remind the 
algorithmic approach of~\cite{mm_infigFI} and then we revisit the classical definition. 
From this, we shall infer the new approach explained in Section~\ref{gros_infig}.

\section{Infinigons and infinigrids: classical approach}
\label{clas_infig}

   We remind the reader that we consider the Poincar\'e's disc model of the hyperbolic plane.
We denote by~$D$ the once and for all fixed disc of the Euclidean plane which is the support
of Poincar\'e's model. We denote by $\partial D$ the circle which is the border of~$D$. 
We remind the reader that the points of~$\partial D$ are called {\bf points at infinity}
and that they do not belong to the hyperbolic plane.
The figures of the paper will take place in this frame. In our sequel, 
otherwise not mentioned, {\it line}
means a line of the hyperbolic plane, most often an arc of a circle in the model.
We refer the reader to~\cite{mmbook1}, for instance, where other references are mentioned.

   In this paper, we give a different proof from what was outlined in~\cite{mm_infigFI}
although it is based on the same construction.

Fix two orthogonal diameters of~$D$. One is called {\bf horizontal} and the other
one {\bf vertical}. In order to define the infinigons, we consider the following
sequence $\{x_n\}_{n\in Z\!\!Z}$. 
Given two points $x_{n}$ and $x_{n+1}$ with $n\in I\!\!N$ and the angle $\alpha\in]0,\pi[$, we first
construct the line $\beta_{n+1}$ which passes through~$x_{n+1}$ and which makes an
angle of $\alpha/2$ with the line $x_nx_{n+1}$. Define
$x_{n+2}$ to be the image of~$x_n$ by reflection in~$\beta_n$. We
repeat the process indefinitely, starting from~\hbox{$x_0=$~O}, where O is the centre of~$D$, 
and from~$x_1$ for
the points with positive indices and starting from~$x_1$ and~$x_0$ for the
points with negative indices. 

\begin{thm} {\rm(Margenstern, see~\cite{mm_infigFI})}\label{const1}
For all $\alpha$ and $x$, the points $x_n$ which
are obtained by the construction above  belong to a euclidean circle,
call it~$\Gamma$ whose diameter is
$\displaystyle{x\over{\cos({\alpha\over2})}}$. 
Moreover, the curvilinear abscissa of the $x_n$'s on $\Gamma$
starting from $x_0$ toward $x_1$ are increasing. 
$\Gamma$ is strictly inside~$D$, is a horocycle or an equidistant curve if and only if 
$x<\displaystyle{\cos({\alpha\over2})}$,
$x=\displaystyle{\cos({\alpha\over2})}$ or
$x>\displaystyle{\cos({\alpha\over2})}$ respectively. If $\Gamma$ is a horocycle, 
the $x_n$'s converge to its unique point at infinity. If $\Gamma$ is an equidistant curve,
the $x_n$'s for positive~$n$ converge to one point at infinity of~$\Gamma$ while
the $x_n$'s for negative~$n$ converge to the other point at infinity.
\end{thm}

   The proof is illustrated by Figure~\ref{construc}. 

\vskip 0pt
\def\boxempty{\hbox{\vbox{\hsize=7pt\offinterlineskip
\ligne{
\vrule height 7pt depth 0pt width 0.6pt
\vbox to 7pt{\hsize=5.8pt
\hrule height 0pt depth 0.6pt width 5.8pt
\vfill
\hrule height 0.6pt depth 0pt width 5.8pt
}\hskip-0.5pt
\vrule height 7pt depth 0pt width 0.6pt
}}
}}

   Let $\mu_1$ be the hyperbolic bisector of the segment $[x_0x_1]$. Then let
$\beta_1$ be the hyperbolic line passing through~$x_1$ which makes
an angle of $\displaystyle{\alpha\over2}$ with the line which supports
$x_0x_1$. Note that we may consider that the first two points are on a diameter of~$D$ with 
$x_0$ at the centre of~$D$. Depending on the possible intersection of~$\mu_1$ with~$\beta_1$,
we have three cases: either $\mu_1$ and~$\beta_1$ meet inside~$D$, or they meet on~$\partial D$
or they do not meet at all.

   First case: $\mu_1\cap\beta_1=A$, where $A$ is a point of the hyperbolic plane, so that
it is inside~$D$, not on~$\partial D$. Then, the triangle $x_0x_1A$ is isosceles with~$x_0x_1$ as 
its basis. The reflection of the triangle in~$\beta_1$ defines a new triangle~$x_1x_2A$ and,
from the reflection, the bisector~$\mu_2$ of~$x_1x_2$ passes through~$A$ and the image~$\beta_2$
of~$x_0A$ is the line~$\beta_2$ which passes through~$x_2$ and which makes an angle 
of~$\displaystyle{\alpha\over2}$  with the hyperbolic line~$x_1x_2$: in particular, $\beta_1$
appears to be the bisector of the angle $(x_1x_0,x_1x_2)$. It is easy to see that
repeating with~$\beta_2$ and the triangle~$x_1x_2A$ what was performed with~$\beta_1$ and
the triangle~$x_0x_1A$, we get a sequence~$x_n$ where the bisector~$\beta_{n+1}$ of the angle 
$(x_{n+1}x_n,x_{n+1}x_{n+2})$ and the bisector~$\mu_{n+1}$ of the segment $[x_nx_{n+1}]$
meet all at the point~$A$. From this, we conclude that all the $x_n$'s belong to a 
circle~$\Gamma$ which, in the Poincar\'e's disc model is a Euclidean circle inside~$D$
with no intersection with~$\partial D$.

   Second case: $\mu_1\cap\beta_1=$~{\bf P}, where {\bf P} is on $\partial D$. This means
that the lines $\mu_1$ and~$\beta_1$ are parallel. The triangle $x_0x_1${\bf P} is not
an ordinary triangle, but what we call an {\bf ideal} triangle as it has one vertex
on~$\partial D$. It is also an isosceles triangle
as \hbox{$(x_1x_0,x_1\hbox{\bf P})=(x_0x_1,x_0\hbox{\bf P})$} by parallelism: this comes from
the fact that $\mu_1$
is the bisector of~$[x_0x_1]$. As the reflection in a line keeps the angles, keeps the
distance and as the images of parallel lines are also parallel lines under a reflection
in a line, We can repeat the argument for the first case and conclude that
all bisectors~$\beta_n$ and~$\mu_n$ constructed with the sequence of the~$x_n$'s meet
at~{\bf P}. Accordingly, the~$x_n$'s are all on a horocycle~$\Gamma$ which, in the Poincar\'e's 
disc model appears as a Euclidean circle which is tangent to~$\partial D$ at~{\bf P}.
  
   Third case: $\mu_1\cap\beta_1=\emptyset$. This time the line~$\mu_1$ and~$\beta_1$ have
a unique common perpendicular~$\pi$. Consider the orthogonal projections $y_0$ and~$y_1$ of 
$x_0$ and~$x_1$ respectively on~$\pi$. By the reflection in~$\mu_1$, $x_0y_0y_1x_1$ is a Saccheri
quadrangle. Now, $\beta_1=x_1y_1$ so that the reflection in~$\beta_1$ which keeps~$\pi$ globally
invariant provides us with a new Saccheri quadrangle $x_2y_2y_1x_1$. In this new setting,
it clearly appears that $x_1y_1$ is the bisector of the angle $(x_1x_0,x_1x_2)$. Repeating
the process, we have two sequences, the $x_n$'s and the $y_n$'s, their orthogonal projection
on~$\pi$. We can see that the bisectors of the segments $x_nx_{n+1}$ and the bisectors of the
angles $(x_{n+1}x_n,x_{n+1}x_{n+2})$ are all perpendicular to~$\pi$. Moreover, as
$x_{n+1}y_{n+1}y_nx_n$ is a Saccheri quadrangle, it is clear that all lengths $x_ny_n$
are equal so that the $x_n$'s lie on an equidistant line~$\Gamma$ which is an Euclidean circle
in the Poincar\'e's disc model and this time, $\Gamma$ precisely has two points of intersection
with $\partial D$.

     Denote by $\vert x_0x_n\vert_\Gamma$ the length of $x_0x_n$, in the hyperbolic plane, 
taken on~$\Gamma$. In the last two cases, \hbox{$\vert x_0x_n\vert_\Gamma=n\vert x_0x_1\vert$}.
In the first case, if $C_\Gamma$ is the circumference of~$\Gamma$, we have that
\hbox{$\vert x_0x_n\vert_\Gamma=n\vert x_0x_1\vert$ mod $C_\Gamma$}. We can characterize
the condition on the Euclidean length ~$x_0x_1$, when $x_0=0$, for which $\Gamma$ is either
a circle, a horocycle or an equidistant curve.

    As we assume that $x_0=0$, it is not difficult to see that $x_0x_1$ is supported by
a diameter of~$D$ and that the hyperbolic line passing through~$x_0$ and making an
angle of $\displaystyle{\alpha\over2}$ with $x_0x_1$ is also a diameter of~$D$. Accordingly,
the Euclidean diameter of~$\Gamma$ is $\displaystyle{{x_0x_1}\over{\cos({\alpha\over2})}}$. Now,
it is clear that $\Gamma$ intersects $\partial D$ if and only if
$\displaystyle{{x_0x_1}\over{\cos({\alpha\over2})}}\geq 1$.

   In the case of a horocycle, as \hbox{$\vert x_nx_{n+1}\vert_\Gamma=\vert x_0x_1\vert_\Gamma$}
for all~$n$, the sequence cannot converge to a point which would be strictly inside~$D$.
As \hbox{$\vert x_0x_n\vert_\Gamma=n\vert x_0x_1\vert$} and as the $x_n$'s
have an accumulation point in the closure of~$D$ which is compact, the $x_n$'s converge
to the unique point at infinity of~$\Gamma$. The same argument shows the conclusion of the
theorem when $\Gamma$ is an equidistant curve. 
\hfill \boxempty

\vtop{
\ligne{\hfill\includegraphics[scale=0.7]{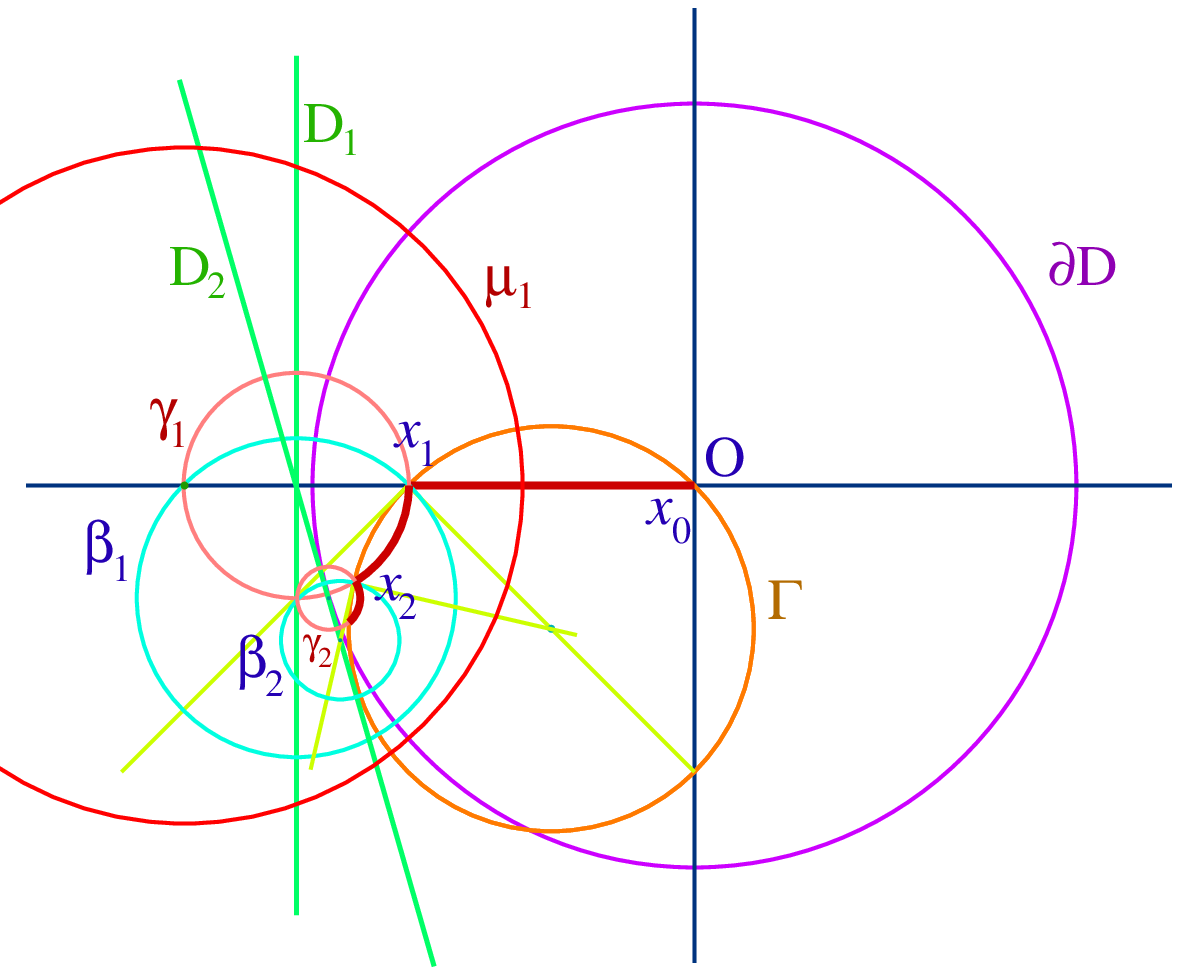}
\hfill}
\begin{fig}\label{construc}
\leurre The construction described in the proof of Theorem~{\rm\ref{const1}}.
\end{fig}
}

   In the case when $x=\displaystyle{\cos({\alpha\over2})}$, the convex hull of the $x_n$'s,
completed by the reflection in one of the $\beta_n$'s is called an {\bf infinigon}. It is not
difficult to see that an infinigon does not occupy all the hyperbolic plane and that it tiles
the plane if and only if $\displaystyle{\alpha\over2}=\displaystyle{{2\pi}\over k}$ for some
$k$ with $k\geq 3$.

    However, there is another construction of the infinigons which will be of help for us.
It was also indicated in~\cite{mm_infigFI}. It consists in considering a regular convex
polygon~$P$ with 
$\displaystyle{\pi\over2}$ as interior angle, by placing a vertex~$V$ at~O, the centre of~$D$,
and one side abutting at~$V$ on a diameter of~$D$, and the other side on a diameter which is
orthogonal to the previous one. Next, with these conditions being fixed, we make the number of
sides of~$P$ to grow. What happens? It happens the length of the side of~$P$ increases, but
it reaches a limit, namely $\displaystyle{\cos({\pi\over4})}$ in this example. And so, the limit
of the polygons is an infinigon as we have indicated. This can be generalized as follows:

\begin{thm}\label{const2}
Let $P_{p,q}$ denote the regular convex polygon such that one of its vertex is~{\rm O}. 
Let $\Gamma_{p,q}$ be the circumscribed circle of~$P_{p,q}$. We may assume that the tangent 
at~{\rm O} to~$\Gamma_{p,q}$ is horizontal. Let $h_{p,q}$ be the Euclidean distance of the
hyperbolic centre of~$\Gamma_{p,q}$ to~{\rm O} and let $e_{p,q}$ be the Euclidean distance of the
Euclidean centre of~$\Gamma_{p,q}$ to~{\rm O}. Then,
\vskip 5pt
\ligne{\hfill
$h_{p,q}=\displaystyle{{\cos({\pi\over q}+{\pi\over p})}%
\over{\sqrt{\cos^2{\pi\over q}-\sin^2{\pi\over p}}}}$. 
\hfill $(1)$\hskip 20pt}
\vskip 5pt
\noindent
and:
\vskip 5pt
\ligne{\hfill
$e_{p,q}=\displaystyle{{\cos({\pi\over q}+{\pi\over p})\sqrt{\cos^2{\pi\over q}-\sin^2{\pi\over p}}}
\over{\cos^2{\pi\over q}-\sin^2{\pi\over p}+\cos^2({\pi\over q}+{\pi\over p})}}$
\hfill $(2)$\hskip 20pt}
\vskip 5pt
We have that for any~$q$, $h_{p,q}\rightarrow 1$ as $p\rightarrow\infty$ and
$e_{p,q}\rightarrow \displaystyle{1\over2}$ as $p\rightarrow\infty$.
\end{thm}

\def\vect#1{\setbox215=\hbox{$#1$}
               \vbox{\parindent0pt\leftskip 0pt\hsize=\wd215
                     \advance\hsize by 5.6pt
                     \ligne{\rightarrowfill}
                     \vskip-6pt
                     \ligne{\hfill\box215\hfill}
                    }
               }

Let us consider this situation. Let $P_{p,q}$ be the regular convex polygon with $p$ sides
and an interior angle~$\displaystyle{{2\pi}\over q}$ so that $q$~copies of $P_{p,q}$
can be put around a point~$A$ to cover a neighbourhood of~$A$ with no overlap.
As previously, we can put one vertex
of the polygon at~O and then, we proceed as in the proof of the theorem. As we know, the $x_n$'s
are on a circle~$\Gamma_{p,q}$. In~\cite{mmbook3}, we give the computation of the radius 
of~$\Gamma_{p,q}$. Here we give a somewhat simplified version of this computation.
The vertices of~$P_{p,q}$ can be written as $r_{p,q}e^{i\vartheta}$ which we rewrite 
$re^{i\vartheta}$ as $p$ and~$q$ are fixed in this part of the proof. 
We define $\vartheta$ by $\vartheta=\displaystyle{-{\pi\over p}+k\displaystyle{{2\pi}\over p}}$ with $k\in[0..p$$-$$1]$. Define $A_k$ the vertex defined by $k$ and consider $A_0$. The Euclidean
support~$C$ of the segment $A_0A_1$ is the circle whose equation is
\hbox{$X^2+Y^2-2\omega X+1=0$}, where $(\omega,0)$ is the centre~$\Omega$ of~$C$. 
Let $(x_0,y_0)$ be the coordinates of~$A_0$. We write that the tangent of~$C$ at~$A_0$ makes
the angle $\displaystyle{\pi\over q}$ with $OA_0$. As $\vect{\Omega A_0}$ has coordinates
$(x_0-\omega,y0)$, we can take $\vect T$ with coordinates $(y_0,\omega-x_0)$ for the tangent.
As 
\hbox{$\vect{\hbox{\rm O}A_0}.\vect T = %
\hbox{\rm O}A_0.\vert \vect T\vert.\cos\displaystyle{\pi\over q}$}, this equality gives us:
\hbox{$y_0\omega=\sqrt{x_0^2+y_0^2}\sqrt{(\omega-x_0)^2+y_0^2}\cos\displaystyle{\pi\over q}$}. As
$y_0\not=0$, we divide this by~$y_0$ giving us:
\vskip 5pt
\ligne{\hfill
$\omega=\displaystyle{\sqrt{1+{{x_0^2}\over{y_0^2}}}\sqrt{\omega^2-1}\cos{{\pi}\over q}}$
\hfill $(a)$\hskip 20pt}
\vskip 3pt
\noindent
In $(a)$, we can see that \hbox{$\displaystyle{{x_0}\over{y_0}}=%
\displaystyle{{\cos{{\pi}\over p}}\over{\sin{{\pi}\over p}}}$} and from the equation of~$C$ which
passes through~$A_0$, we get that \hbox{$(\omega-x_0)^2+y_0^2=\omega^2-1$}, so that squaring~$(a)$
and putting in it the just obtained equalities we obtain after easy simplifications:
\ligne{\hfill
$\omega^2=\displaystyle{{\cos^2{{\pi}\over q}}%
\over{\cos^2{{\pi}\over q}-\sin^2{{\pi}\over p}}}$
\hfill $(b)$\hskip 20pt}
\vskip 3pt
\noindent
Now, we rewrite the fact that $C$ passes through~$A_0$ by
\hbox{$r^2-2\omega r\cos\displaystyle{{\pi}\over p}+1=0$}, solving this equation in~$r$ and
taking into account that \hbox{$0<r<1$} should be satisfied, we get:
\vskip 5pt
\ligne{\hfill
$r=\omega\cos\displaystyle{{\pi}\over p}%
-\displaystyle{\sqrt{\omega^2\cos^2{{\pi}\over p}-1}}$. 
\hfill $(c)$\hskip 20pt}
\vskip 3pt
\noindent
From~$(c)$ we easily get that \hbox{$r=\displaystyle{{{\cos{{\pi}\over q}\cos{{\pi}\over p}}%
\over{\sqrt\Delta}}-{{\sqrt N}\over{\sqrt\Delta}}}$}, where 
$\Delta=\displaystyle{\cos^2{{\pi}\over q}-\sin^2{{\pi}\over p}}$
and \hbox{$N=\displaystyle{\cos^2{{\pi}\over q}\cos^2{{\pi}\over p}-\cos^2{{\pi}\over q}%
+\sin^2{{\pi}\over p}}$}. Factorizing $\displaystyle{\cos^2{{\pi}\over q}}$ in $N$ and then
$\displaystyle{\sin^2{{\pi}\over p}}$ we get that
\hbox{$N=\displaystyle{\sin^2{{\pi}\over q}\sin^2{{\pi}\over p}}$} which allows us to
easily deduce~$(1)$.

   In order to get the distance from the Euclidean centre of~$\Gamma_{p,q}$ to~O, let $h=h_{p,q}$
in order to simplify the notations and let $s$ be the Euclidean length of the diameter
of~$\Gamma_{p,q}$. Of course, the required distance is $\displaystyle{s\over2}$, so that we have
to compute~$s$. Now, let $S$ be the point at distance~$s$ from~O on a diameter of~$D$ which also
passes through the centre of~$\Gamma_{p,q}$ and let $H$ denote the hyperbolic centre
of~$\Gamma_{p,q}$. We know that $H$ is the hyperbolic mid-point of O$S$. Let $C$ be the
circle which passes through~$H$, centred on the line O$S$ and which is orthogonal to~$\partial D$.
Let $(\omega,0)$ be the coordinates of~$C$ as O$S$ can be taken as the $x$-axis. 
We have:
\vskip 5pt
\ligne{\hfill
$\omega(\omega-s)=(\omega-h)^2$\hfill}
\ligne{\hfill
$h^2-2h\omega+1=0$\hfill}
\vskip 5pt
The first equation says that $H$ is the mid-point of O$S$. The second one says that $C$ passes
through~$H$. Cancelling $\omega^2$ in the first equation and subtracting with the second one
we get that \hbox{$\omega s = 1$}. Putting that in the second equation we get
\hbox{$h^2s-2h+s=0$} from which we derive $s=\displaystyle{{2h}\over{1+h^2}}$. Using $(1)$
in this latter expression, we get $(2)$ after straightforward computations.

The conclusion of the Theorem on the limits of $h_{p,q}$ and $e_{p,q}$ are straightforward.
\hfill\boxempty

We can see that the conclusion of Theorem~\ref{const2} confirms the statement of 
Theorem~\ref{const1}.

\section{Infinigons, infinigrids and grossone}
\label{gros_infig}

 What can be said about these construction in the light of the new numeral system introduced
by Yaroslav Sergeyev, see~\cite{sergeyev1,sergeyev2,sergeyev3,sergeyev4,sergeyev5}? Here, We mainly look at the case when
$x=\displaystyle{\cos({\alpha\over2})}$ and when \hbox{$\alpha=\displaystyle{{2\pi}\over q}$}
for some positive integer~$q$. 

   Let us consider the construction involved in Theorem~\ref{const1}.
We can see that we must replace the vague notion of $\infty$ with the more
precise indication on the infinite number of $x_n$'s we consider. The conclusion we must
reach is that if $\lambda$ is a positive infinite integer, $x_\lambda$ never reaches {\bf P}. 
Indeed, by 
construction, $x_\lambda$ is in the hyperbolic plane, {\it i.e.} inside~$D$, so that if 
$x_{\lambda+1}$ can be defined, $x_{\lambda+1}$ is also in the hyperbolic plane and it cannot 
be~{\bf P}.
However, {\bf P} itself can be defined, at least as the intersection of~$\beta_1$ and~$\mu_1$
at infinity or, which is equivalent, by saying that $\beta_1$ and~$\mu_1$ are parallel.
Accordingly, we can say that in some sense, the infinigon can also be defined, but it is
an {\bf ideal} object in this sense that it is essentially incomplete: we cannot tell the number
of its sides.
 
   If we look at the construction which is considered in Theorem~\ref{const2}, we have
a completely different landscape. This time, if $\lambda$ is a positive infinite integer,
we can define a regular convex polygon~$P$ with $\lambda$~sides. The computations performed
in Section~\ref{clas_infig} for Theorem~\ref{const2} gives us a precise description of this
object. Consider that $q$~is fixed, where $q$~is a positive finite integer. Then
the vertices of~$P$ are on a circle~$\Gamma$ of the hyperbolic plane which is not a horocycle:
this circle is completely in the hyperbolic plane. However, its diameter is infinite as
its representation in the hyperbolic plane is infinitesimally close to~1. Replacing~$p$
by~$\lambda$ in~$(2)$, we can see that the diameter of~$\Gamma$ is
\hbox{$\displaystyle{{2\cos({{\pi}\over q}+{{\pi}\over \lambda})%
\sqrt{\cos^2{{\pi}\over q}-\sin^2{{\pi}\over \lambda}}}%
\over{\cos^2{{\pi}\over q}-\sin^2{{\pi}\over \lambda}+%
\cos^2({{\pi}\over q}+{{\pi}\over \lambda})}}$}.
Let us put \hbox{$a=\displaystyle{\cos({{\pi}\over q}+{{\pi}\over \lambda})}$}
and $b=\displaystyle{\cos^2{{\pi}\over q}-\sin^2{{\pi}\over \lambda}}$. Then,
\hbox{$d=\displaystyle{{2a\sqrt b}\over{a^2+b}}$}. Is it true that $d<1$? Indeed,
$d<1$ if and only if \hbox{$\displaystyle{2a\sqrt b < a^2+b}$} {\it i.e.}
if and only if \hbox{$4a^2b < a^4+b^2+2a^2b$} which is equivalent to
\hbox{$a^4+b^2-2a^2b>0$}. It remains to see that \hbox{$a^2\not=b$}.
Indeed:

\ligne{\hskip 20pt$b-a^2=%
\displaystyle{\cos^2{{\pi}\over q}-\sin^2{{\pi}\over \lambda}-\cos^2({{\pi}\over q}+%
{{\pi}\over \lambda})}$
\hfill}
\ligne{\hskip 50pt
$=\displaystyle{\cos^2{{\pi}\over q}-\sin^2{{\pi}\over \lambda}-\cos^2{{\pi}\over q}\cos^2{{\pi}%
\over \lambda}%
-\sin^2{{\pi}\over q}\sin^2{{\pi}\over \lambda}}$\hfill}
\ligne{\hskip 92.5pt
$+~2\displaystyle{\cos{{\pi}\over q}\cos{{\pi}\over \lambda}\sin{{\pi}\over q}%
\sin{{\pi}\over \lambda}}$
\hfill}
\ligne{\hskip 50pt
$=\displaystyle{\cos^2{{\pi}\over q}(1-\cos^2{{\pi}\over \lambda})-\sin^2{{\pi}\over \lambda}
-\sin^2{{\pi}\over q}\sin^2{{\pi}\over \lambda}}$\hfill}
\ligne{\hskip 92.5pt
$+~2\displaystyle{\cos{{\pi}\over q}\cos{{\pi}\over \lambda}\sin{{\pi}\over q}%
\sin{{\pi}\over \lambda}}$
\hfill}
\ligne{\hskip 50pt
$=\displaystyle{\cos^2{{\pi}\over q}\sin^2{{\pi}\over \lambda}-\sin^2{{\pi}\over \lambda}
-\sin^2{{\pi}\over q}\sin^2{{\pi}\over \lambda}}$\hfill}
\ligne{\hskip 92.5pt
$+~2\displaystyle{\cos{{\pi}\over q}\cos{{\pi}\over \lambda}\sin{{\pi}\over q}%
\sin{{\pi}\over \lambda}}$
\hfill}
\ligne{\hskip 50pt
$=\displaystyle{(\cos^2{{\pi}\over q}-1)\sin^2{{\pi}\over \lambda}
-\sin^2{{\pi}\over q}\sin^2{{\pi}\over \lambda}}$\hfill}
\ligne{\hskip 92.5pt
$+~2\displaystyle{\cos{{\pi}\over q}\cos{{\pi}\over \lambda}\sin{{\pi}\over q}%
\sin{{\pi}\over \lambda}}$
\hfill}
\ligne{\hskip 50pt
$=\displaystyle{-2\sin^2{{\pi}\over q}\sin^2{{\pi}\over \lambda}
+2\cos{{\pi}\over q}\cos{{\pi}\over \lambda}\sin{{\pi}\over q}\sin{{\pi}\over \lambda}}$
\hfill}
\ligne{\hskip 50pt
$=\displaystyle{2\sin{{\pi}\over q}\sin{{\pi}\over \lambda}
(\cos{{\pi}\over q}\cos{{\pi}\over \lambda}-\sin{{\pi}\over q}\sin{{\pi}\over \lambda})}$
\hfill}
\ligne{\hskip 50pt
$=\displaystyle{2\sin{{\pi}\over q}\sin{{\pi}\over \lambda}
\cos({{\pi}\over q}+{{\pi}\over \lambda})}>0$
\hfill}
\vskip 5pt
\noindent
because \hbox{$\displaystyle{\sin{{\pi}\over q}}>0$}, 
\hbox{$\displaystyle{\sin{{\pi}\over \lambda}}>0$}
and \hbox{$\displaystyle{\cos({{\pi}\over q}+{{\pi}\over \lambda})}>0$}
as \hbox{$\displaystyle{0<{{\pi}\over q}+{{\pi}\over \lambda}< {{\pi}\over 2}}$} which is true
since the isosceles triangles which constitute~$P$ are true triangles 
in the hyperbolic plane. 

\vtop{
\ligne{\hfill\includegraphics[scale=0.625]{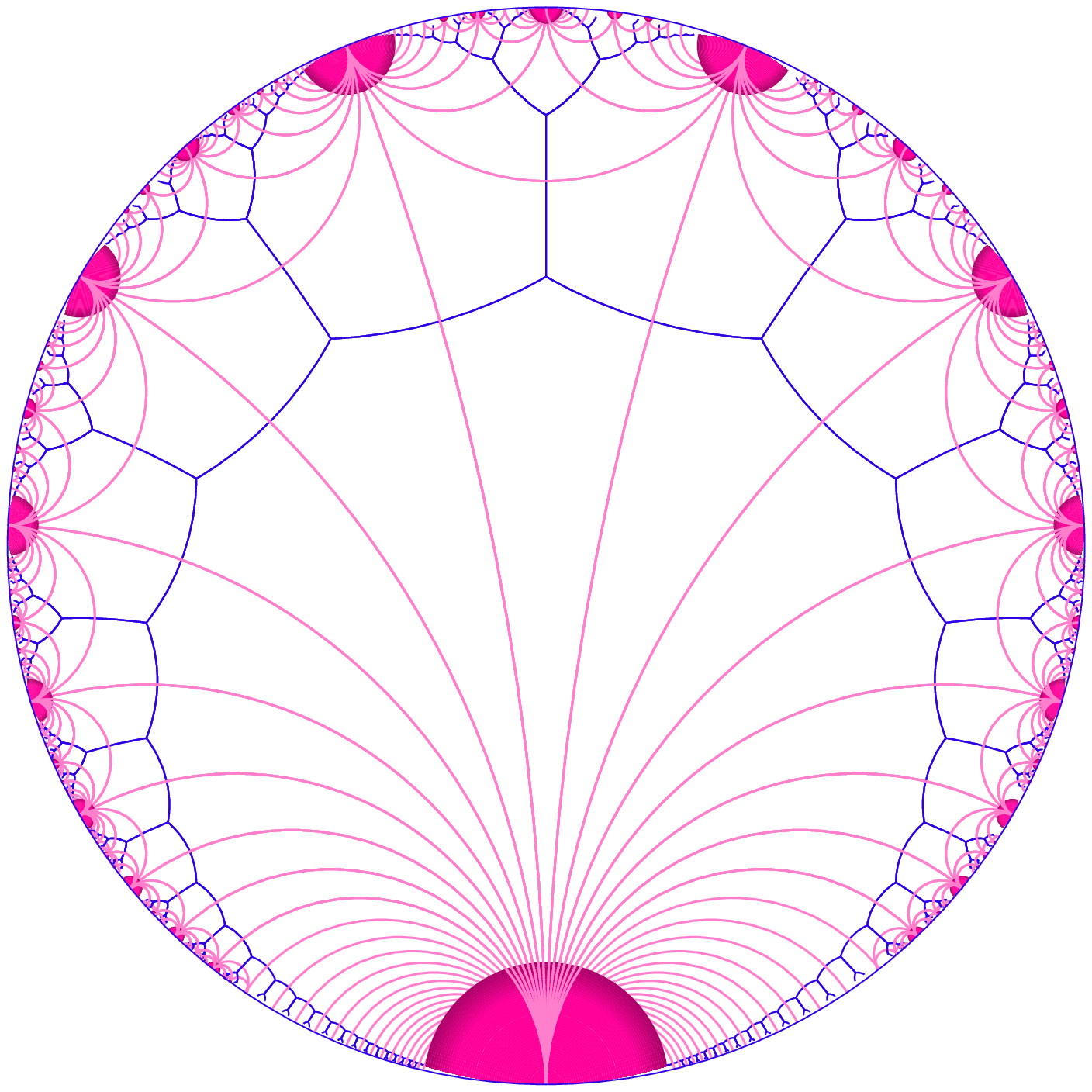}
\hfill}
\begin{fig}\label{infigon}
\leurre An illustration for the first order infinigons. In this picture, $q=3$.
\end{fig}
}

So we proved that:
\vskip 5pt
\ligne{\hfill
$\displaystyle{b-a^2= 2\sin{{\pi}\over q}\sin{{\pi}\over \lambda}
\cos({{\pi}\over q}+{{\pi}\over \lambda})}$
\hfill $(3)$\hskip 20pt}
\vskip 3pt
   Accordingly, we have proved that $d<1$ which shows that $\Gamma$ and~$P$ both remain in the
hyperbolic plane.

   The just provided computation can be made more precise: we know
that \hbox{$\displaystyle{(a^2+b)^2-(2a\sqrt b)^2 = (a^2-b)^2}$} so that from~$(3)$
as $d=2h_{p,\lambda}$, we can see that 
\hbox{$\displaystyle{1-d^2=({{b-a^2}\over{a^2+b}})^2}$}. When $\lambda$
is an infinite positive integer, the order of \hbox{$a^2+b$} is 
\hbox{$2\displaystyle{\cos^2{{\pi}\over q}}$} and that of $b-a^2$ is 
\hbox{$2\displaystyle{\sin{{\pi}\over q}\cos{{\pi}\over q}}$}, so that
\vskip 5pt
\ligne{\hfill
$d^2\approx\displaystyle{1-%
{{\sin^2{{\pi}\over q}}\over{\cos^2{{\pi}\over q}}}{{\pi^2}\over{\lambda^2}}}$
\hfill $(4)$\hskip 20pt}.
\vskip 3pt
This inequality proves that $\Gamma$ is still in the hyperbolic plane.
Call {\bf infinigon} this polygon with infinitely many sides, exactly with $\lambda$~of them.
It is plain that such an infinigon tiles the hyperbolic plane.

   In fact, from~$(1)$ and from the fact that the hyperbolic distance~$\delta$ corresponding to the
Euclidean distance~$d$ from~O to~$S$ is given by Lobachevsky's formula:
\hbox{$\displaystyle{\delta=\ln\Big\vert{{1+d}\over{1-d}}\Big\vert}$}. We obtain in this way
that $\delta=\displaystyle{2\ln\Big({{a+\sqrt b}\over{\sqrt b-a}}\Big)}$ so that in the end
$\delta=2\ln(a+\sqrt b)-2\ln(\sqrt b-a)$. Now,:
\vskip 5pt
\ligne{\hskip 20pt 
$\ln(a+\sqrt b)=2\ln\Big(\displaystyle{\cos({\pi\over q}}%
+{\pi\over\lambda})+\sqrt{\cos^2{\pi\over q}-\sin^2{\pi\over\lambda}}\Big)$ 
\hfill}
\vskip 5pt
\noindent
which is
equal to $2\ln\Big(\cos(\displaystyle{\pi\over q}+\displaystyle{A\over\lambda})\Big)$, 
where $A$ is a function of~$q$
and~$\lambda$ bounded by a finite positive number. On another hand,
\vskip 5pt
\ligne{\hskip 20pt 
$\ln(\sqrt b-a)=2\ln\Big(
\displaystyle{\sqrt{\cos^2{\pi\over q}-\sin^2{\pi\over\lambda}}}
-\displaystyle{\cos({\pi\over q}}%
+{\pi\over\lambda})
\Big)$ 
\hfill}
\ligne{\hskip 70pt 
$=2\ln\Big(
\displaystyle{\sqrt{\cos^2{\pi\over q}-\sin^2{\pi\over\lambda}}}
-\displaystyle{\cos({\pi\over q}}%
+{\pi\over\lambda})
\Big)$ 
\hfill}
\vskip 5pt
\noindent
Now, $\displaystyle{\sqrt{\cos^2{\pi\over q}-\sin^2{\pi\over\lambda}}}
=\displaystyle{\cos{\pi\over q}\sqrt{1-\tan^2{\pi\over\lambda}}}
\approx\displaystyle{(1-{{\pi^2}\over{2\lambda^2}})\cos{\pi\over q}}$, as 
$\displaystyle{\tan{\pi\over\lambda}}$ differs from $\displaystyle{\pi\over\lambda}$
by a higher order infinitesimal. Taking this into account, the previous computation
can be continued as follows:
\vskip 5pt
\ligne{\hskip 20pt 
$\ln(\sqrt b-a)=2\ln\Big(\displaystyle{(1-{{\pi^2}\over{\lambda^2}})\cos{\pi\over q}}
-\displaystyle{\cos{\pi\over q}}\cos{\pi\over\lambda}+
\sin{\pi\over q}\sin{\pi\over\lambda}
\Big)$ 
\hfill}
\ligne{\hskip 70pt 
$=2\ln\Big(\displaystyle{(1-\cos{\pi\over \lambda})\cos{\pi\over q}}
-{{\pi^2}\over{\lambda^2}}\cos{\pi\over q}+\sin{\pi\over q}\sin{\pi\over\lambda}
\Big)$
\hfill}
\ligne{\hskip 70pt 
$=2\ln\Big(\displaystyle{2\sin({\pi\over {2\lambda}})^2\cos{\pi\over q}}
-{{\pi^2}\over{\lambda^2}}\cos{\pi\over q}+\sin{\pi\over q}\sin{\pi\over\lambda}
\Big)$
\hfill}
\ligne{\hskip 70pt 
$\approx 2\ln\Big(\displaystyle{{{\pi^2}\over {2\lambda^2}}\cos{\pi\over q}}
-{{\pi^2}\over{\lambda^2}}\cos{\pi\over q}+{\pi\over\lambda}\sin{\pi\over q}
\Big)$.
\hfill}

As $\displaystyle{1\over\lambda}$ is an infinitesimal which is infinitely bigger
than $\displaystyle{1\over{\lambda}^2}$, we get that

\ligne{\hskip 20pt
$\ln(\sqrt b-a)\approx 2\ln\Big(\displaystyle{{\pi\over\lambda}\sin{\pi\over q}}\Big)=
-2\ln\lambda+2\ln\pi+2\ln\displaystyle{\sin{\pi\over q}}$. 
\hfill}

As $\delta=2\ln(a+\sqrt b)-2\ln(\sqrt b-a)$, we eventually get
that $\delta\approx 2\ln\lambda$, so that $\delta$ is an infinite number.

    Now, from~$(1)$ and~$(2)$ we have something more: if we replace~$q$ by a positive infinite
integer~$\mu$, the formulas are still valid as well as the computation leading to formula~$(3)$.
In this case we can replace the estimation given in~$(4)$ by the following one:
\vskip 5pt
\ligne{\hfill
$d^2\approx1-\displaystyle{{{\pi^4}\over{\lambda^2\mu^2}}}$
and $\delta\approx 2\ln\displaystyle{{\pi^2}\over{\lambda\mu}}
\approx -2\ln\lambda-2\ln\mu$.
\hfill $(5)$\hskip 20pt}
\vskip 3pt

    Again we call {\bf infinigon} the polygon obtained in this case as its diameter
is actually infinite and as the number of its sides is also defined by an infinite integer. 
However, in order
to distinguish between these two kinds of infinigons, call {\bf infinigon of first order},
for short {\bf first order infinigon},
those defined by $P_{q,\lambda}$ where $\lambda$ is a positive infinite integer and $q$~a positive
finite integer with $q\geq 3$. We call {\bf infinigon of second order}, for short
{\bf second order infinigon}, those defined 
by $P_{\mu,\lambda}$ where $\lambda$ and~$_mu$ are both positive infinite integers.
It is plain that both $P_{q,\lambda}$ and $P_{\mu,\lambda}$ tile the plane by the standard
process: we take $P_0$ a copy of~$P_{q,\lambda}$ and then we replicate it by reflection in its
sides and, recursively, by reflections of the images in their sides. The same process can be applied
to copies of~$P_{\mu,\lambda}$ where both $\mu$ and~$\lambda$ are infinite numbers.

   We have still one point to investigate.

   When we say that the infinigons of first order tile the plane by the above process, we say 
{\it recursively} which is in fact a vague term. In the traditional meaning, this means 
{\it endlessly}. As there is no more precise notion than the cardinals for estimating infinite 
numbers in traditional mathematics, here we have to make things more clear. When we say 
{\it recursively} we have to mention to which depth we go on the recursive process. Controlling
recursion up to a fixed depth in advance is a standard feature in the implementation of
certain programming languages. This does not prevent more theoretic oriented languages to allow
depths which are only limited by the resources of the machine on which the program runs.
Here, we adopt the same spirit: when we use the word {\it recursively}, it is possible
to not indicate to which depth, but for a precise study of the process, it is better to
indicate to which depth we allow to proceed. 
Let~$\nu$ be the depth of recursion. It is plain that from this definition, after a few steps
of iteration of the process, we may obtain a copy which overlaps an already existing copy. Of course,a copy is considered to be reached by the $k^{\rm th}$ recursive call if it has not been produced
by a previous call. 

A way to detect the set of copies obtained up to the depth~$n$ has been indicated 
in~\cite{mm_infigFI}. It consists in building a tree which is in bijection with the
tiling. However, as~\cite{mm_infigFI} was written with a more traditional look at infinity,
we have to revisit this construction.

Say that the centre of an infinigon is the centre of its circumscribing circle.
We fix two contiguous sides of~$P_0$, a fixed copy of~$P_{q,\lambda}$, say $s_0$ and~$s_1$
and let~$V_0$ be their common vertex. 
We have two cases depending on whether $q$~is odd or even.

First assume that $q$~is even, say $q=2h$. This is the easy case. Consider the ray~$\ell$ which 
is issued from~$V_0$ and which supports~$s_0$. We may consider that $s_0$ lies on the left hand 
side of~$V_0$. Let~$V_0^1$ be the 
other end of~$s_0$. From~$V_0^1$, out of~$s_0$, $\ell$ is the support of a side shared by two 
copies of~$P_0$ which share~$V_0^1$ with~$P_0$. It is plain that this process can be continued.
The same can be performed with~$m$, the ray issued from~$V_0$ which supports~$s_1$. Let~$\cal S$
be the angular sector defined by the angle $(\ell,m)$. We construct a tree whose root is
attached to~$P_0$.
Each node of the tree is attached to a copy of~$P_0$ inside~$\cal S$. A node~$\nu$ of the tree
is the son of a node~$\pi$ only if the copy attached to~$\nu$ and that attached to~$\pi$
share a common side. To precisely define the notion of childhood, we start from the root.
By definition, its sons are the reflections of~$P_0$ in its sides which are still inside~$\cal S$.
Now, consider~$\sigma_k$, $k\in[1..\lambda-2]$, the other sides of~$P_0$, starting from~$V_0^1$ 
and counter-clockwise turning around~$P_0$. Let $P_k$ be the reflection of~$P_0$ in~$\sigma_k$.
Let~$V_0^k$ be the vertex shared by~$\sigma_{k-1}$ and~$\sigma_k$ with $\sigma_0=s_0$. Then,
around~$V_0^k$, there are $q$~copies of~$P_0$, $P_0$ being taken into account. If we recursively
repeat the process of copying starting from the $P_k$'s, there will be overlapping: starting from
$P_1$ and looking at its sides on the right-hand side, after $q$$-$2 iterations, we get $P_2$.
Accordingly, we have to give rules in order to avoid overlapping. To this purpose, we shall say
that the $P_k$'s we have just defined with $k\in[1..\lambda-2]$ are the {\bf main sons}.
Each main son have $j$~{\bf brothers}, $j$~being $q$$-$3 or $q$$-$4, depending on a circumstance
to which we turn now.

    Consider $\sigma_k$, a side of~$P_0$ in the angle~$(\ell,m)$ with $k>1$ and $k<\lambda$$-$2.
The vertices of~$\sigma_k$ are $V_0^k$ and~$V_0^{k-1}$. In order to delimit regions which do not
overlap but completely cover the complement of~$P_0$ in the angle~$(\ell,m)$, we continue
$\sigma_k$ a,d $\sigma_{k+1}$ both to the left. Now, it is easy to see that around $V_0^k$ and
outside $P_0$, we have $h$~copies of the angle~$(\ell,m)$, so that outside $P_k$ and
around~$V_0^k$ there are $h$$-$1 copies of~$(\ell,m)$. On the other side, there are
$h$$-$2 copies of~$(\ell,m)$ as from the continuation of~$\sigma_{k+1}$ we have to take into
account the angle which is in~$P_0$ and that which is in~$P_k$. So that for 
\hbox{$1<k<\lambda$$-$2}, in each region $(\sigma_k,\sigma_{k+1})$ and outside~$P_k$
$q$$-$3 copies of~$(\ell,m)$. We remain with the examination of $k=1$
and $k=\lambda$$-$2. When $k=\lambda$$-$2, $s_1$ plays the role of~$\sigma_{k+1}$, so that 
in this case, the number of angles left in the region $(\sigma_{\lambda-2},s_1)$, is also
$q$$-$3. When $k=1$, The region is~$(s_0,\sigma_1)$ which is smaller than a region 
$(\sigma_k,\sigma_{k+1})$ with $k>1$. The region is smaller than~$\pi$ by an angle which
is equal to~$(\ell,m)$, so that this time we have $h$$-$2 copies of $(\ell,m)$ close to~$V_0~1$
and $h$$-$2 of them to close to~$V_0^2$, so that we get $q$$-$4 copies of~$(\ell,m)$.
Accordingly, in this case, we can see that it is possible to split $(\ell,m)$ into
\hbox{$1+\lambda$$-$$2$} copies of~$P_0$ and \hbox{$(\lambda$$-$$3)(q$$-$$3)+q$$-$$4$} copies
of~$(\ell,m)_2$ where $(\ell,m)_2$ has the same angle as $(\ell,m)$ but with a depth
reduced by~2: if $P_0$ has recursion depth~$\kappa$, with $\kappa$ finite or infinite integer,
$P_k$ with $k\in[1..\lambda$$-$$2]$ has depth~$\kappa$+1, so that any infinigon in
an $(\ell,m)_2$ has depth at least $\kappa$+2. Note that $h$$-$1 steps of recursion are needed
in order the vertices of~$P_0$ should be completely covered.

    Now, consider the case when $q$~is odd, we shall write $q=2h$+1.

This time, the regions have to be changed. Consider the previous setting with the
same notations. If we continue the side~$\sigma_k$, it is not a side of an adjacent
infinigon: the continuation is a bisector of an interior angle of an infinigon which shares $V_0^k$ with $\sigma_k$.
Now, to consider half-infinigons is possible only if $\lambda$ is even. If $\lambda$~too is odd,
then we get a more complex situation. 

However, using a trick we explained in~\cite{mmarXiv_newpq}, we can handle the situation no matter which the parity of~$\lambda$ is. The idea consists in replacing the regions we considered when $q$~is even 
by new regions to which we now turn. Consider the mid-point$M_0$ of the side~$s_0$ of~$P_0$. Let $N_0$ be the mid-point of the side~$\tau_0$ which abuts~$s_0$ at~$V_0^1$ and which makes an angle $\vartheta=\displaystyle{h{\pi\over q}}$
with~$s_0$ by going outside of~$P_0$. Let~$W_0$ be the other end of~$\tau_0$.
Let $R_0$ be the mid-point of the side~$\omega_0$ which abuts~$\tau_0$ at~$W_0$
and which makes an angle $\varphi_0=\displaystyle{h{\pi\over q}}$ with~$\tau_0$, the angles $\vartheta_0$ and~$\varphi_0$ being on different sides of~$\tau_0$.
From the construction, the isosceles triangles $M_0V_0^1N_0$ and
$N_0W_0R_0$ are equal so that the points $M_0$, $N_0$ and $R_0$ lie on a same ray~$u$ issued from~$M_0$ whose supporting line is called a {\bf $h$-mid-point line} in~\cite{mmarXiv_newpq}. A similar
$h$-mid-point line~$v$ can be drawn from the mid-point of~$s_1$ which is symmetric to~$u$ with respect to the bisector $\beta$ of the angle $(\ell,m)$. Clearly, $u$ and~$v$ meet on$\beta$ inside~$P_0$. Such an $(u,v)$ is called an {\bf angular  sector} $(u,v)$ and we may distinguish between copies of it depending on the recursion depth of the copy.
In such an angular sector $(u,v)$, outside~$P_0$, we take all infinigons such that all mid-points of their sides lie inside the angle $(u,v)$ or, possibly, on~$u$ or on~$v$. 

       Using the construction of the rays~$u$ and~$v$, we can define a process which is
similar to the one we defined for the case when $q$~is even. This time, for each $\sigma_k$ with \hbox{$1<k<\lambda$$-$2}, we consider the rays issued from
the mid-points of~$\sigma_{k-1}$ and~$\sigma_{k+1}$ supported by $h$-mid-point lines with respect to~$\sigma_{k-1}$ on one side and with respect to~$\sigma_{k+1}$ on the other side. This define new regions which we call {\bf strips}. All these strips are equal and, indeed, the equality also holds for $k=1$ and 
\hbox{$k=\lambda$$-$1}. Besides $P_k$, each strip
contains $q$$-$4 copies of an angular sectors $(u,v)$ with a smaller depth, smaller by~2.

Note that the sides of infinigons which cross $u$ and~$v$ can be used to define the depth of the 
recursion. Consider an angular sector $(u,v)$ as defined above. After$s_0$, denote by $s_i$ the 
sides of infinigons which cross the $h$-mid-point line which supports~$u$. We have that $s_{2i}$ 
goes inside the sector while $s_{2i+1}$ goes outside. The infinigon whose side is $s_{j+1}$ is 
reached from that whose side is~$s_j$ after $h$$-$1 reflections. It is more natural to count the 
recursion depth in this way so that after one recursion step, the vertices of the previous 
generation of infinigons are completely covered by the new one.  We shall now take this definition 
of the recursion depth which we also call generation.

 Accordingly, if $N_{k+1}$ is the number of infinigons generated at the 
$k+1^{\rm th}$ generation, then, from what we have proved we can see that
$N_{k+1}=(\lambda$$-$$2)(q$$-$$4)N_k$ when $q$ is odd and
$N_{k+1}=((\lambda$$-$$2)(q$$-$$3)$$-$$1)N_k$ when $q$ is even as:
\vskip 3pt
\ligne{\hfill
$(\lambda$$-$$3)(q$$-$$3)$+$q$$-$$4=(\lambda$$-$$2)(q$$-$$3)$$-$1.
\hfill}

We can sum up our study by the following result:

\begin{thm}
There are two kinds of regular convex infinigons in the hyperbolic plane: those which have $\lambda$
sides and $\displaystyle{{2\pi}\over q}$ as their interior angle, where $\lambda$ is an infinite 
integer and $q$ is a finite one, and those which have $\lambda$ sides and 
$\displaystyle{{2\pi}\over \mu}$ as their interior angle, where both $\lambda$ and~$\mu$ are 
infinite integers. The first kind of infinigons are said of the {\bf first order} and the second 
kind are said of {\bf second order}. Both first order and second order infinigons lie completely 
in the hyperbolic plane with no point at infinity: in both cases, there is a hyperbolic circle, 
whose radius is infinite, which circumscribes all the vertices. The radius $\rho_{\lambda,p}$,
$\rho_{\lambda,\mu}$ of these circles, when passing through~{\rm O}, the centre of the 
Poincar\'e's disc is given by the following formulas:
\vskip 3pt
\ligne{\hfill
$\rho_{\lambda,q}=\displaystyle{{\cos({\pi\over q}+{\pi\over \lambda})}%
\over{\sqrt{\cos^2{\pi\over q}-\sin^2{\pi\over \lambda}}}}$, and  
$\rho_{\lambda,\mu}=\displaystyle{{\cos({\pi\over \mu}+{\pi\over \lambda})}%
\over{\sqrt{\cos^2{\pi\over \mu}-\sin^2{\pi\over \lambda}}}}$.
\hfill}
\vskip 3pt
\noindent
where $\lambda$ and $\mu$ are infinite integers, the left-, right-hand side formula applying to 
first, second order infinigons respectively. Both kinds of infinigons tile the hyperbolic plane, 
giving rise to two kinds of infinite families of tilings: 
${\cal T}_{\lambda,q,\nu}$ and ${\cal T}_{\lambda,\mu,\nu}$, where $\nu$ is an infinite positive 
integer, indicating the depth of the recursion used to define the tiling.
For first order infinigons, the number of tiles in  ${\cal T}_{\lambda,q,\nu}$
is $((\lambda$$-$$2)(q$$-$$4))^\nu$ when $q$ is odd,
and it is $((\lambda$$-$$2)(q$$-$$3)-1)^\nu$ when $q$ is even. For second order infinigons, the 
number of tiles in ${\cal T}_{\lambda,\mu,\nu}$
is $(\lambda$$-$$2)(\mu$$-$$4))^\nu$ when $\mu$ is odd and it is
$((\lambda$$-$$2)(\mu$$-$$3)-1)^\nu$ when $\mu$ is even.
\end{thm}
   
   Note that as $q$$-$4 is $2h$$-$3 when $q=2h+1$ and $q$$-$3 is $2h$$-$3 too when $q=2h$.




\begin{thebibliography}{5}
\bibitem{mmarXiv_newpq}
M. Margenstern,
About a new splitting for the algorithmic study of the tilings $\{p,q\}$ of the hyperbolic plane when $q$ is odd, {\it arXiv}:0911.4040v2, (2009), 20pp.
\end{thebibliography}
\end{document}